# ANOMALOUS RESISTANCE IN CRITICAL IONIZATION VELOCITY PHENOMENA

V.I. Badin

*Institute of Terrestrial Magnetism, Ionosphere and Radiowave Propagation, Troitsk, Moscow Region, 142190, Russia*

## ABSTRACT

To describe the generation of the electric field by a discontinuity of the Hall current, an equation of the third order is obtained using the electric charge conservation and Ohm laws. The solutions of this equation are used to model the electric impulses detected in experiments aimed to verify the Alfven hypothesis on the critical ionization velocity at collisions of neutral gas with magnetized plasma. A quantitative agreement with experiment is attained and the main features of measured signals are modeled under an assumption on the strong anomalous resistance behind the discontinuity. Apparently, the anomalous resistance occurs due to trapping the current carriers by a small-scaled modulation of the electric field.

INTRODUCTION

H. Alfven (1960) first proposed the ionization of the neutral gas streaming across magnetic field lines to explain some features of the Solar System. According to his hypothesis, a gas can be ionized when its velocity exceeds the critical value $V_c=(2e\varphi_i/M)^{1/2}$ where $\varphi_I$ is the ionization potential of the atom, $M$ is its mass, and $e$ is the absolute value of the electron charge, with the gas velocity being determined in the reference frame where the transversal electric field is zero. This hypothesis encouraged wide experimental studies (Brenning, 1992). First, there were laboratory experiments with homopolars and plasma guns. Later, space experiments with injections of rapidly ionized vapors were carried out The number of space experiments executed is now rather great, though some phenomena remain not completely understood yet. The generation of electric fields and precipitation of energetic particles are among these phenomena. It is more convenient to study the physical mechanisms of these processes in a laboratory experiment since the latter can provide simultaneous measurements of all necessary parameters such as particle number densities and temperatures, electric and magnetic fields.

Strong electric fields (or electric structures) produced by the interchanging positive and negative volume charges were detected in laboratory and space experiments (Piel, Mobius, and Himmel, 1980; Venkataramani and Mattoo, 1986; Kelley et al., 1991). The electric fields exhibit a form of potential barriers or wells with amplitudes of the order of hundred Volts. These structures named as ionization fronts were supposed to play a key role in the ionization process. For today, there is no generally accepted theory of the generation of such structures, although there is a qualitative model (Piel, Mobius, and Himmel, 1980). N. Brenning (1992) noticed that a discontinuity of the Hall current could be among the possible mechanisms for generating the strong electric fields.

V. Badin (1997, 1998) proposed to model the generation of these structures in terms of the theory of the dynamo-effect in a strong discontinuity taking into account the time changes in the volume density of electric charges. One of the objectives of this work is to elaborate the proposed model of the dynamo-effect in order to analyze laboratory experiments.

FORMULATING THE PROBLEM

Consider an ionized gas moving along the *x*-axis of the rectangular Cartesian coordinate system. Direct the *z*-axis along the magnetic field. Take a reference frame in which the component $E_y$ of the electric field equals zero. Assume a strong discontinuity moving along the *x*-axis with the velocity $u_0$. Assume also the characteristic scale of the studied structure along the *y*-axis is much greater than that along the *x*-axis and neglect the *y*-derivatives. Use the equations for the density $\rho$ of electric charges

$$\partial\rho/\partial t + \text{div}\, \boldsymbol{j} = 0, \qquad (1)$$

$$\text{div}\, \boldsymbol{E} = 4\pi\rho. \qquad (2)$$

Assume that the current density obeys the Ohm law $\boldsymbol{j} = \sigma\boldsymbol{E}$, with the conductivity $\sigma$ being different for two sides of the discontinuity. Then we obtain from Eqs. (1) and (2)

$$\partial^2 E_x/\partial t\partial x + 4\pi\sigma_P \partial E_x/\partial x + 4\pi\partial j_z/\partial z = 0, \qquad (3)$$

where $\sigma_P$ is the Pedersen conductivity and $j_z$ is the current along the magnetic field.

In laboratory experiments, the longitudinal conductivity was high enough and the longitudinal field was not observed (Brenning, 1992). Therefore, we can integrate Eq. (3) by *z* supposing $\partial E_x/\partial z \equiv 0$. Assume that the field is symmetrical with respect to the line $z = 0$, and the walls of the discharge volume are situated at $z = \pm L$ where *L* is a half of the height of the discharge tube along the magnetic field. Upon integrating from *0* to *L*, the Eq. (3) contains the current $j_z(L)$ to the wall of the discharge tube. As a rule, the wall is made from an insulating material (glass) which contains conducting elements (probes, other instruments etc.). A metal cover on the outer surface shields the experimental device. Assume that the electric field is potential $E = -\text{grad } \varphi$ and the shield potential equals zero. For this case, the current to the wall can be estimated as

$$j_z(L) = \Gamma \partial \varphi / \partial t + \varphi / R, \qquad (4)$$

where $\Gamma$ is the average electric capacitance and $R$ is the average active resistance of the wall of the discharge tube per unit area.

Let us find a solution of the Eq. (3) which moves with the discontinuity i.e. depends only on $\xi = x - u_0 t$. For this case, the Eqs. (3) and (4) yield

$$d^3\varphi/d\xi^3 - K_P d^2\varphi/d\xi^2 - K_C k d\varphi/d\xi + K_R^2 k \varphi = 0, \qquad (5)$$

where $K_P = 4\pi\sigma_P/u_0$, $K_C = 4\pi\Gamma$, $K_R^2 = 4\pi/Ru_0$, and $k = 1/L$.

Let the subscript "1" denote the gas behind the discontinuity ($\xi < 0$) and the subscript "2" correspond to the gas before the discontinuity ($\xi > 0$). The boundary conditions can be written as follows (Badin, 1997):

$$\sigma_{P1} \partial \varphi / \partial \xi - \sigma_{P2} \partial \varphi_2 / \partial \xi = (\sigma_{H1} u_1 - \sigma_{H2} u_2) B/c \equiv I \qquad \text{at } \xi = 0, \qquad (6)$$

where $\sigma_H$ is the Hall conductivity, $B$ is the magnetic field, $c$ is the speed of light, and $I$ is a jump of the Hall current;

$$\varphi_1 = \varphi_2 \qquad \text{at } \xi = 0; \qquad (7)$$

$$\int_{-\infty}^{+\infty} j_z(L) d\xi = IL \qquad (8)$$

(assume $I = \text{const}$ for simplicity).

ELECTRIC SHOCK IMPULSE

We can express the solution of the Eq. (5) (tending to zero at $\xi \to \pm\infty$) in the following form:

$$\varphi_1 = C_1 \exp(\lambda_1 \xi) + C_3 \exp(\lambda_3 \xi) \qquad \text{for } \xi < 0, \qquad (9)$$

$$\varphi_2 = C_2 \exp(\lambda_2 \xi) \qquad \text{for } \xi > 0, \qquad (10)$$

where the constants $C_i$ are determined by the boundary conditions, and the numbers $\lambda_i$ are roots of the characteristic equation

$$\lambda^3 - K_P \lambda^2 - K_C k \lambda + K_R^2 k = 0, \qquad (11)$$

with $\text{Re}(\lambda_1, \lambda_3) > 0$ and $\lambda_2 < 0$. The discriminant of the Eq. (11) is proportional to the expression

$$Q = 1 - 4K_P^3/27kK_R^2 - 2K_C K_P/3K_R^2 - K_C^2 K_P^2/27K_R^4 - 4kK_C^3/27K_R^4. \qquad (12)$$

For $Q > 0$, the dimensionless ratios in the Eq. (12) are obviously restricted. To simplify calculations, we assume them small enough. Note that the smallness of the parameter $\eta = K_P^3/kK_R^2$ indicates a small Pedersen conductivity in comparison with the conductivity of the experimental device. For this case, the complex conjugate roots of the Eq. (11) approximately are

$$\lambda_{1,3} \approx (kK_R^2)^{1/3}(1/2 \pm i3^{1/2}/2).$$

For a high Pedersen conductivity, when $\eta \gg 1$ and $Q < 0$, the real negative root of the Eq. (11) approximately is

$$\lambda_2 \approx -(kK_R^2/K_P)^{1/2} - kK_C/2K_P.$$

For the obtained roots $\lambda_i$, we can approximately express the solution of the Eq. (5) satisfying the boundary conditions (6–8) in the following form:

$$\varphi_1 \approx \frac{2I}{\sqrt{3}\Delta_0} \exp(\tfrac{1}{2}l_1^{-1}\xi)[\frac{l_1}{R_1}\sin(\tfrac{\sqrt{3}}{2}l_1^{-1}\xi + \pi/3) - \frac{K_c}{R_2 K_{R2}^2}\sin(\tfrac{\sqrt{3}}{2}l_1^{-1}\xi)] \qquad \text{for } \xi < 0, \tag{13}$$

$$\varphi_2 \approx \frac{Il_1}{\Delta_0 R_1}\exp(-l_2^{-1}\xi) \qquad \text{for } \xi > 0, \tag{14}$$

where the lengths

$$l_1 = (kK_{R1}^2)^{-1/3} \text{ and } l_2 = (kK_{R2}^2/K_{P2})^{-1/2} \tag{15}$$

are characteristic scales of the electric shock impulse behind the discontinuity and in front of it respectively, and

$$\Delta_0 = \frac{\sigma_{P1}}{R_2}\frac{l_2}{l_1} + \frac{\sigma_{P2}}{R_1}\frac{l_1}{l_2}. \tag{16}$$

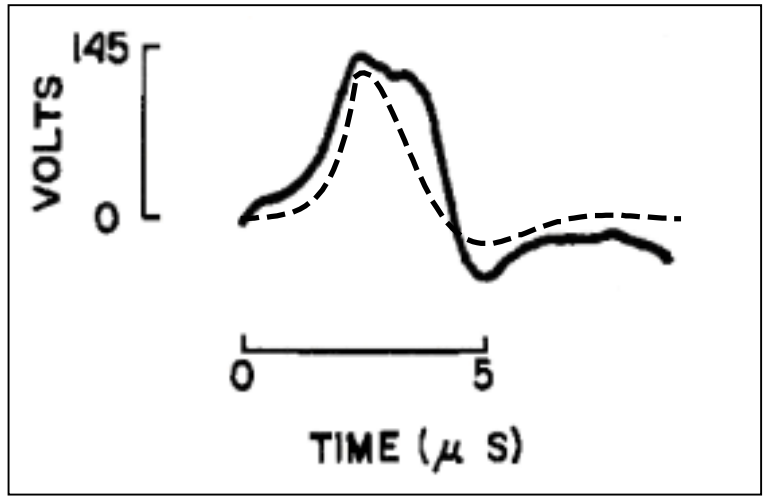

Fig. 1. The floating probe potential (solid curve) and electric shock calculated by the formulae (13, 14) (dashed curve).

COMPARISON WITH EXPERIMENT

Figure 1 presents the floating probe potential (solid curve) in the experiment with a plasma gun (Venkataramani and Mattoo, 1986). The potential profile corresponds to the neutral number density (argon atoms) $N_n \sim 10^{13}$ cm$^{-3}$ and electron number density $N_e \sim 5 \cdot 10^{12}$ cm$^{-3}$. The model of conductivity of a weakly ionized gas yields the following nominal estimates: $\sigma_P \approx 2.6 \cdot 10^{10}$ s$^{-1}$ and $\sigma_H \approx 1 \cdot 10^{10}$ s$^{-1}$. The plot shows that the oscillations of potential were detected only on the trailing edge of the impulse. In the framework of the proposed model, this indicates that the leading edge of the impulse is determined by the real (negative) root of the characteristic equation while the complex conjugate roots correspond to the trailing edge.

To calculate the scales of the impulse, one should estimate the active resistance $R$. It is the most probable that the current to the wall takes its path through the measuring circuits and/or other conducting elements of the experimental device. In this case, the current closes along the wall of the discharge tube across the magnetic field lines. Consequently, we can estimate the resistance as $R \sim d/\sigma_P$ where $d$ is a characteristic scale length. Assume that $d \sim 1$ cm (the distance between the Langmuir probes was about 2 cm).

The experimental potential profile is well approximated, if we assume $l_1 = 7$ cm (experimenters indicated the impulse scale of 5 cm). The formulae (15) yield for this case $l_2 \approx 4.5$ cm and $\sigma_{P1} \approx 4.6 \cdot 10^4 \text{s}^{-1}$. Assuming for the $\sigma_{P2}$ its nominal value and supposing that the jump of the Hall current is of the order of the nominal current itself $\sigma_H B u_0 /c$, we obtain by the formulae (13, 14) the potential profile which is shown in Figure 1 by the dashed curve. We can see that the measured and calculated curves agree, if we assume for $\sigma_{P1}$ the value indicated above that differs from nominal by a factor exceeding $10^5$. Note that such conductivities can really provide $\eta_1 \ll 1$ and $\eta_2 \gg 1$ in accordance with our assumptions.

The proposed model gives an explanation for the experimentally discovered features of the observed electric structures: the scaling that is a relation of the width of potential barrier to the neutral gas density and the invariance of the potential maximum with respect to this density (Venkataramani and Mattoo, 1986). Namely, the Pedersen conductivity of a rarified gas is proportional to the number density of neutrals and, according to the formula (15), the scale length $l_1$ is proportional to $N_n^{-1/3}$ that agrees with experiments. For low neutral gas densities, the maximum of the potential (13) approximately equals
$\varphi_{1max} \approx I u_0 \Gamma \exp(-\pi/3^{1.5})/\Delta_0$.
This quantity is almost independent of the neutral gas density since the Hall conductivity of a rarefied gas is proportional to $N_n^2$.

It is unlikely that a strong anomalous resistance behind the discontinuity can be associated with scattering in the plasma turbulence since the inertia of heavy ions is rather strong. If we supposed that the anomalous resistance is attributed to the growth in the effective collision frequency, this frequency would exceed $2 \cdot 10^{10}$ s$^{-1}$ i.e. exceed the Larmor frequency of electrons. Such strong scattering would provide an isotropic plasma conductivity and a longitudinal electric field that was not observed in the experiment.

Apparently, trapping the current carriers (first of all ions) by a small-scaled modulation of the electric field is responsible for this anomalous resistance. Scattering can diminish the energy of the directed motion of ions and the ions can be trapped. The small-scaled modulation of the electric field is clearly seen in experimental potential profiles for low neutral gas densities (Venkataramani and Mattoo, 1986).

## DISCUSSION

A hypothesis on the possible strong plasma resistance involves into equations the time derivative of the volume density of electric charges. This increases the order of the modeling equation and brings into consideration dissipative structures as modeling objects. In contrast to waves, these structures are completely localized in the vicinity of their generation that complicates observations.

The use of a dissipative structure as a model for the detected electric impulse enables us to attain an agreement with experiments. The proposed model can explain the main features of measured signals under an assumption on the very strong resistance behind the discontinuity. Apparently, this anomalous resistance is attributed to trapping the current carriers by a small-scaled modulation of the trailing edge of the electric shock.

The proposed mechanism of the anomalous resistance is of interest for the phenomena of acceleration of charged particles. Namely, when it operates in rarified plasma, the trapped particles spatially arranged can produce a strong self-consistent electric field while free particles can gain energy with a weak scattering. This sort of acceleration mechanism is very promising for various resistive processes in space plasmas. Perhaps, it will find technological applications as well.

## ACKNOWLEDGMENT

This work was supported by the Russian Foundation for Basic Research, project no. 99-05-65080.